%%%%%%%%%%%%%%%%%%%%%%%%%%%%%%%%%%%%%%%%%%%%%%%%%%%%%%%%%%%%%%%%%%%%%%%%%%%%%%
%%%%%%%%%%%%%%%%%%%%%%%%%%%%%%%%%%%%%%%%%%%%%%%%%%%%%%%%%%%%%%%%%%%%%%%%%%%%%

%%%%%%%%%%%%%%%%%%%%%%%%%%%%%%%%%%%%%%%%%%%%%%%%%%%%%%%%%%%%%%%%%%%%%%%%%%%%%%%%%%%%%%%%%%%
\input harvmac
% VERSION Feb 10, 2000 
%\input psfig
\newcount\figno
\figno=0
\def\fig#1#2#3{
\par\begingroup\parindent=0pt\leftskip=1cm\rightskip=1cm\parindent=0pt
\global\advance\figno by 1
\midinsert
\epsfxsize=#3
\centerline{\epsfbox{#2}}
\vskip 12pt
{\bf Fig. \the\figno:} #1\par
\endinsert\endgroup\par
}
\def\figlabel#1{\xdef#1{\the\figno}}
\def\encadremath#1{\vbox{\hrule\hbox{\vrule\kern8pt\vbox{\kern8pt
\hbox{$\displaystyle #1$}\kern8pt}
\kern8pt\vrule}\hrule}}

\overfullrule=0pt

%macros
%
\def\underarrow#1{\vbox{\ialign{##\crcr$\hfil\displaystyle
 {#1}\hfil$\crcr\noalign{\kern1pt\nointerlineskip}$\longrightarrow$\crcr}}}
% use of underarrow
%A~~~\underarrow{a}~~~B
%
\def\tilde{\widetilde}

\def\inbar{\vrule height1.5ex width.4pt depth0pt}
\def\IC{\relax\hbox{\kern.25em$\inbar\kern-.3em{\rm C}$}}
\def\IR{\relax\hbox{\kern.25em$\inbar\kern-.3em{\rm R}$}}
\def\IZ{\relax\ifmmode\hbox{Z\kern-.4em Z}\else{Z\kern-.4em Z}\fi}

\font\zfont = cmss10 %scaled \magstep1

\def\bigone{\hbox{1\kern -.23em {\rm l}}}
\def\ZZ{\hbox{\zfont Z\kern-.4emZ}}

%%%%%%%%%%%%%%%%%%%%%%%%%%%%%%%%%%%%%%%%%%%%%%%

%  draw box of size #1pt and line thickness #2pt
\def\drawbox#1#2{\hrule height#2pt
        \hbox{\vrule width#2pt height#1pt \kern#1pt
              \vrule width#2pt}
              \hrule height#2pt}
% Young tableaux

\def\Asym#1#2{\vcenter{\vbox{\drawbox{#1}{#2}
              \kern-#2pt       % line up boxes
              \drawbox{#1}{#2}}}}

\batchmode
  \font\bbbfont=msbm10
\errorstopmode
\newif\ifamsf\amsftrue
\ifx\bbbfont\nullfont
  \amsffalse
\fi
\ifamsf
\def\IR{\hbox{\bbbfont R}}
\def\IC{\hbox{\bbbfont C}}

\def\IZ{\hbox{\bbbfont Z}}

%%%%%%%%%%%%%%%%%%%%%%%%%%%%%%%%%%%%%%%%%%%%%%%%
%\def\fund{\vcenter{\vbox{\drawbox}}}
%\def\antifund{{\ov \fund}}
%\def\fund{n}

\midinsert
\endinsert

%%%%%%%%%%%%%%%%%%%%%%%%%%%%%%%%%%%%%%%%%%%%%%%%%%%%%%%%%%%%%%%%%%%%%%%%%%%%%%%

\nref\bayen{F. Bayen, M. Flato, C. Fronsdal, A. Lichnerowicz and D.
Sternheimer, {\it Ann. Phys.} {\bf 111}, 61 (1978); {\it Ann. Phys.} {\bf 111}, 111
(1978).}

\nref\wilde{M. De Wilde and P.B.A. Lecomte, {\it Lett. Math. Phys.} {\bf 7}, 487
(1983).}

\nref\omory{H. Omori, Y. Maeda and A. Yoshioka, {\it Adv. Math.} {\bf 85}, 
224 (1991).}

\nref\fedosov{ B. Fedosov, J.
{\it Diff. Geom.} {\bf 40}, 213 (1994); {\it Deformation Quantization and Index 
Theory} (Akademie Verlag, Berlin, 1996).}
 
\nref\k{M. Kontsevich, ``Deformation Quantization of Poisson
Manifolds I'', q-alg/9709040; {\it Lett. Math. Phys.} {\bf 48}, 35 (1999).}

\nref\sw{N. Seiberg and E. Witten, ``String Theory and Noncommutative
Geometry'', hep-th/9908142.}

\nref\dito{J. Dito, {\it Lett. Math. Phys.} {\bf 20}, 125 (1990).}

\nref\ditodos{ J. Dito, {\it Lett. Math. Phys.} {\bf 27}, 73 (1993).}

\nref\antonu{ F. Antonsen, {\it Phys. Rev.}  {\bf D56}, 920 (1997).}

\nref\antond{F. Antonsen, ``Deformation Quantization of Constrained
Systems'', gr-qc/9710021.}

\nref\antont{F. Antonsen, ``Deformation Quantization of Gravity'',
gr-qc/9712012.}

\nref\zachos1{ T. Curtright and C. Zachos, {\it J. Phys.}  {\bf A32},
771 (1999).}

\nref\weyl{H. Weyl, {\it Group Theory and Quantum Mechanics}, (Dover, New
York, 1931).}

\nref\kasper{P. Kasperkovitz and M. Peev, {\it Ann. Phys.} {\bf 230}, 21
(1994).}

\nref\pptt{J.F. Pleba\'nski, M. Przanowski, J. Tosiek and F.J.
Turrubiates, ``Remarks on Deformation Quantization on the Cylinder'', to
be published in {\it Acta Phys. Pol.} {\bf B} (1999).}

\nref\strato{R.L. Stratonovich, {\it Sov. Phys. JETP} {\bf 31}, 1012 (1956).}

\nref\graciau{J.M. Gracia Bond\'{\i}a and J.C. Varilly, {\it J. Phys. A: Math.
Gen.} {\bf 21}, L879 (1988).}

\nref\graciau{J.M. Gracia Bond\'{\i}a and J.C. Varilly, {\it Ann. Phys.}
{\bf 190}, 107 (1989).}

\nref\carinena{J.F. Cari\~nena, J.M. Gracia Bond\'{\i}a and J.C. Varilly,
{\it J. Phys. A: Math. Gen.} {\bf 23}, 901 (1990).}

\nref\gadella{M. Gadella, M.A. Mart\'{\i}n, L.M. Nieto and M.A. del Olmo,
{\it J. Math. Phys.} {\bf 32}, 1182 (1991).}

\nref\ppt{ J.F. Pleba\'nski, M. Przanowski and J. Tosiek, {\it Acta Phys. Pol.} 
{\bf B27} 1961 (1996).}

\nref\gross{A. Grossmann, {\it Commun. Math. Phys.} {\bf 48}, 191 (1976).}

\nref\moyal{J.E. Moyal, {\it Proc. Camb. Phil. Soc.} {\bf 45}, 99 (1949).}

\nref\wigner{E.P. Wigner, {\it Phys. Rev.} {\bf 40}, 749 (1932).}

\nref\tata{W.I. Tatarskii, {\it Usp. Fiz. Nauk} {\bf 139}, 587 (1983).}

\nref\hillery{M. Hillery, R.F. O'Connell, M.O. Scully and E.P. Wigner,
{\it Phys. Rep.} {\bf 106}, 121 (1984).}

\nref\bogo{N.N. Bogoliubov and D.V. Shirkov, {\it Introduction to the
Theory of Quantized Fields}, (John Wiley $\&$ Sons, New York, 1980).}

\nref\iz{C. Itzykson and J.B. Zuber, {\it Quantum Field Theory}, (Mc
Graw-Hill, New York, 1980).}

\nref\hatfield{B. Hatfield, {\it Quantum Field Theory of Point Particles
and Strings}, (Addison-Wesley Publishing, Redwood City, 1992).}

\nref\wein{S. Weinberg, {\it The Quantum Theory of Fields} Vol. I, 
(Cambridge University Press, Cambridge, 1995).}

\nref\pita{ V.B. Berestetskii, E.M. Lifshitz and L.P. Pitaevskii,
{\it Relativistic Quantum Theory}, (Pergamon Press, New York, 1971). }

\nref\landau{L.D. Landau and E.M. Lifshitz, {\it The Classical Field
Theory}, (Pergamon Press, New York, 1975).}

\nref\curtright{T. Curtright D. Fairlie and C. Zachos, {\it Phys. Rev.}  
{\bf D58},  025002 (1998).}

\nref\antonc{F. Antonsen, ``Zeta-Functions and Star-Products'',
quant-ph/9802031.}

\nref\aw{G.S. Agarwal and E. Wolf, {\it Phys. Rev.}  {\bf D2} (1970) 2161;
2206.}

\nref\campos{H. Garc\'{\i}a-Compe\'an, J.F. Pleba\'nski, M. Przanowski
and F.J. Turrubiates, ``Deformation Quantization of String Theory'', to appear.}

\nref\key{B.S. Kay, {\it Phys. Rev.} {\bf D20}, 3052 (1979).}

\nref\birrel{ N.D. Birrell and P.C.W. Davis, {\it Quantum Fields in Curved Space}
(Cambridge University Press, Cambridge, 1982).}

\nref\tablas{I.S. Gradshteyn and I.M. Ryzhik, {\it Table of Integrals, Series, 
and Products} (Academic Press, New York, 1980).}

\nref\isham{C.J. Isham, {\it Proc. R. Soc. Lond.} {\bf A362}, 383 (1978).}

%%%%%%%%%%%%%%%%%%%%%%%%%%%%%%%%%%%%%%%%%%%%%%%%%%%%%%%%%%%%%%%%%%%%%%%%%%%%%%

%%%%%%%%%%%%%%%%%%%%%%%%%%%%%%%%%%%%%%%%%%%%%%%%%%%%%%%%%%%%%%%%%%%%%%%%%%%%%%

\Title{hep-th/9909206, CINVESTAV-FIS-99/59}
{\vbox{\centerline{Deformation Quantization of Classical Fields}
\medskip
\centerline{}}}
\smallskip
\centerline{H.
Garc\'{\i}a-Compe\'an,$^{a}$\foot{compean@fis.cinvestav.mx} J.F. 
Pleba\'nski,$^a$\foot{pleban@fis.cinvestav.mx} M.
Przanowski$^{b,a}$\foot{przan@fis.cinvestav.mx} and
F.J. Turrubiates$^a$\foot{fturrub@fis.cinvestav.mx} }
\smallskip
\centerline{\it $^a$Departamento de F\'{\i}sica}
\centerline{\it  Centro de Investigaci\'on y de Estudios Avanzados del
IPN}
\centerline{\it Apdo. Postal 14-740, 07000, M\'exico D.F., M\'exico}
\smallskip
\centerline{\it $^b$Institute of Physics}
\centerline{\it Technical University of \L \'od\'z}
\centerline{\it W\'olcza\'nska 219, 93-005, \L \'od\'z, Poland}
\bigskip
\baselineskip 18pt
\medskip
\vskip 1truecm
\noindent

We study the deformation quantization of scalar and abelian gauge
classical free fields. Stratonovich-Weyl quantizer,
star-products
and Wigner functionals are obtained in field and oscillator
variables. Abelian gauge theory is particularly intriguing 
since Wigner functional is  factorized into a physical part and other one
containing the constraints only. 
Some effects of non-trivial topology within deformation quantization 
formalism are also considered.

%\vskip 1truecm

%PACS numbers: {\it 03.50.-z, 03.50.De, 11.10.-z, 03.65.Db}

\noindent

\Date{September, 1999}

%%%%%%%%%%%%%%%%%%%%%%%%%%%%%%%%%%%%%%%%%%%%%%%%%%%%%%%%%%%%%%%%%%%%%%%%

%%%%%%%%%%%%%%%%%%%%%%%%%%%%%%%%%%%%%%%%%%%%%%%%%%%%%%%%%%%%%%%%%%%%%

\newsec{Introduction}

Recently a great deal of works have been done in deformation 
quantization theory. Deformation quantization was originally proposed
by Bayen et al.$^1$ as 
an alternative to 
the standard procedure of quantization avoiding the more
difficult problem of constructing 
the relevant Hilbert space of the system. In these pioneer works
it was proposed a way of quantizing a classical system 
by deforming the corresponding algebraic structures. Different mathematical 
aspects of deformation quantization were also further explored. Among
them, the  existence proofs of a star-product for any symplectic manifold$^{2-4}$
and also for a Poisson
manifold$^5$. The
latter result has been motivated in part by string theory. A full
treatment of the interplay between string theory and
deformation quantization and, more generally, noncommutative 
geometry was given very recently by Seiberg and Witten$^6$.

Deformation quantization was mainly applied to quantize classical
mechanics. However, it seems to be very interesting to formulate quantum
field theory within deformation quantization program. Recently some works
on this subject have been done$^{7-12}$. 

Our paper is motivated by those works and we attempt to show
systematically how all the deformation quantization formalism
(Stratonovich-Weyl quantizer, Grossmann operator, Weyl correspondence,
Moyal star-product, Wigner function, etc.) can be carry over to quantum
field theory. To this end we use the well known particle interpretation of
free fields. This approach in the case of scalar field was first considered by
Dito$^{7,8}$

In this interpretation many formulas seem to be more natural
and in a sense the deformation quantization in terms of particle
interpretation justifies the respective formulas of the previous 
works$^{9-12}$, given in field variable language.

The paper is organized as follows: In Section 2 we consider deformation
quantization in scalar field theory. Stratonovich-Weyl quantizer,
Grossmann operator, Moyal $*$-product and Wigner functional are defined in
terms of the field variables. In Section 3, the same is given for free
scalar fields with the use of normal coordinates. Wigner functional of the
ground state is found and also Wigner functional for higher states is
considered. Finally, the normal ordering within deformation quantization
program is given. Section 4 is devoted to deformation quantization of
free electromagnetic field. The constraints (Gauss's law) are analyzed in
some detail. Perhaps the most interesting result is that the Wigner
functional can be factorized so that one part contains only the transverse
components of the field and the second part contains the longitudinal
variables and this part vanishes when constraints are not satisfied.
Finally, in Section 5 we consider the Casimir effect in terms of deformation 
quantization formalism. 

This paper is the first one of a series of papers which we are going to
dedicate to deformation quantization of classical fields. The next one
deals with interacting and self-interacting fields.

\vskip 2truecm
%%%%%%%%%%%%%%%%%%%%%%%%%%%%%%%%%%%%%%%%%%%%%%%%%%%%%%%%%%%%%%%%%%%%%
%%%%%%%%%%%%%%%%%%%%%%%%%%%%%%%%%%%%%%%%%%%%%%%%%%%%%%%%%%%%%%%%%%%%%

\newsec{Deformation Quantization in Scalar Field Theory}

Consider a real scalar field on Minkowski space time of signature
$(+,+,+,-).$
Canonical variables of this field will be denoted by $\phi (x,t)$ and
$\varpi (x,t)$ with $(x,t) \in \IR^3 \times \IR$. We deal with fields at
the instant $t=0$ and we put $\phi(x,0) \equiv \phi(x)$ and $\varpi(x,0)
\equiv \varpi(x)$. It is worth to mention that some of the functional
formulas and their manipulations of Sec.2 and 3, will be formal.

\vskip 1truecm
%%%%%%%%%%%%%%%%%%%%
\subsec{The Stratonovich-Weyl Quantizer}
Let $F[\phi,\varpi]$ be a functional on the phase space ${\cal Z}$ and let
$\tilde{F}[\lambda,\mu]$
be its Fourier transform

\eqn\uno{\tilde{F}[\lambda,\mu] = \int {\cal D}\phi {\cal D} \varpi
exp\bigg\{-i \int dx \bigg(
\lambda({x}) \phi({x}) + \mu({x}) \varpi({x}) \bigg) \bigg\}
F[\phi , \varpi ].}

The functional measures are given by ${\cal D}\phi = \prod_{x} d \phi
(x), {\cal D} \varpi = \prod_{x} d \varpi (x)$.
By analogy to quantum mechanics$^{13}$ we define the Weyl quantization
rule as follows

\eqn\dos{\hat{F}= W(F[\phi,\varpi]) := \int {\cal D}({\lambda\over 2 \pi})
{\cal D}
({\mu \over 2 \pi})
\tilde{F}[\lambda,\mu] \hat{\cal U}[\lambda,\mu],}
where $\{ \hat{\cal U}[\lambda,\mu]: (\lambda,\mu) \in {\cal Z}^* \}$
is a family of unitary operators

\eqn\tres{ \hat{\cal U}[\lambda,\mu]:= {\rm exp}\bigg\{i\int dx \bigg(
\lambda({x})
\hat{\phi}({x}) +
\mu({x})
\hat{\varpi}({x})\bigg)\bigg\},}
with $\hat{\phi}$ and $\hat{\varpi}$ being the field operators  given by
$ \hat{\phi}({x}) |\phi\rangle = \phi({x}) |\phi \rangle$
and $\hat{\varpi}({x})
|\varpi\rangle = \varpi({x}) |\varpi \rangle.$

Using the well known Campbell-Baker-Hausdorff formula and the standard commutation rules
for $\hat{\phi}$ and $\hat{\varpi}$
one can write $\hat{\cal U}[\lambda, \mu]$ in the following form

$$ \hat{\cal U}[\lambda,\mu] =  exp\bigg \{ -{i \hbar \over 2}
\int dx \lambda({x}) \mu({x}) \bigg \}
exp\bigg \{ i \int dx \mu({x}) \hat{\varpi}({x}) \bigg \}
exp\bigg\{ i\int dx \lambda ({x}) \hat{\phi}({x})\bigg \} $$

\eqn \siete {= exp\bigg \{{i \hbar \over 2} \int dx \lambda({x})
\mu({x}) \bigg \}
exp\bigg \{ i \int dx \lambda({x}) \hat{\phi}({x}) \bigg \}
exp\bigg \{ i \int dx \mu({x}) \hat{\varpi}({x})\bigg \}.}

Employing Eqs. \siete\ and the relations
$\int {\cal D} \phi | \phi \rangle \langle \phi | = \hat{1}$
and $\int {\cal D} ({\varpi \over 2 \pi \hbar}) |\varpi
\rangle \langle \varpi | = \hat{1}$ we easily get

$$ \hat {\cal U}[\lambda, \mu]
= \int {\cal D} ({\varpi \over 2 \pi \hbar})
exp \bigg \{i \int dx \mu ({x}) \varpi ({x}) \bigg \} | \varpi +
{\hbar \lambda \over 2}
\rangle \langle \varpi - {\hbar \lambda \over 2} | $$

\eqn\diez{
= \int {\cal D} {\phi}
exp \bigg \{i \int dx \lambda ({x}) \phi ({x})\bigg \} | \phi -
{\hbar \mu \over 2}
\rangle \langle \phi + {\hbar \mu \over 2}|. }
This operator satisfies the following properties

\eqn\once{ Tr \bigg \{ \hat {\cal U} [\lambda,\mu] \bigg \} = \int {\cal
D} \phi \langle \phi | \hat{\cal U}[\lambda,\mu] |\phi \rangle = \delta
\big[{\hbar \lambda \over 2 \pi}\big] \delta [\mu] }
and

\eqn\doce{Tr \bigg \{ \hat{\cal U}^{\dagger} [\lambda, \mu] \hat{\cal U}
[\lambda ', \mu '] \bigg \} = \delta [ {\hbar \over 2 \pi}(\lambda -
\lambda ')] \delta [\mu - \mu '].}
(Compare with Refs. 14 and 15).

Substituing \uno\ into \dos\ one has

\eqn\trece{ \hat {F} = W(F[\phi,\varpi])= \int {\cal D} \phi {\cal D} ({\varpi \over 2 \pi
\hbar}) F[\phi, \varpi] \hat{\Omega} [\phi, \varpi],}
where $\hat{\Omega}$ is given by

\eqn\catorce {\hat{\Omega} [\phi, \varpi] = \int {\cal D} ({\hbar \lambda
\over 2 \pi})  {\cal D} \mu exp \bigg \{ -i \int dx \bigg(\lambda (x)
\phi (x) + \mu (x) \varpi (x) \bigg) \bigg \} \hat{\cal U} [\lambda,\mu].}

It is evident that the operator $\hat{\Omega}$ defined by \catorce\  is
the quantum field analog of the well known Stratonovich-Weyl quantizer
playing  an important role in deformation quantization of classical
mechanics$^{16-21}$.
Therefore we call our $\hat{\Omega}$ the Stratonovich-Weyl quantizer.  
One can easily check the following properties of $\hat{\Omega}$
   
\eqn\quince{ \big (\hat{\Omega} [\phi,\varpi] \big)^{^\dagger} =
\big(\hat{\Omega} [\phi,\varpi] \big),}

\eqn\dseis{ Tr \big \{ \hat{\Omega}[\phi , \varpi] \big \} = 1, }

\eqn\dsiete{ Tr \bigg \{ \hat{\Omega}[\phi,\varpi] \hat{\Omega}[\phi
',\varpi '] \bigg \} = \delta [\phi - \phi'] \delta [{\varpi - \varpi '
\over 2 \pi \hbar}]. }

Multipliying \trece\ by $\hat{\Omega}[\phi,\varpi]$, taking the trace of
both sides and using \dsiete\  we get

\eqn\docho{ F[\phi , \varpi] = Tr \bigg \{ \hat{\Omega} [\phi , \varpi]  
\hat{F} \bigg\}. }

One can also express $\hat{\Omega}[\phi,\varpi]$ in a very useful form
by inserting \diez\ into \catorce. Thus one gets

$$ \hat{\Omega}[\phi, \varpi] = \int {\cal D} ({\eta \over 2
\pi
\hbar}) exp \bigg \{ -{i \over \hbar} \int dx \eta(x) \phi (x) \bigg \}
|\varpi + {\eta \over 2} \rangle \langle \varpi - {\eta \over 2}| $$
   
\eqn \dnueve { = \int {\cal D} \xi exp \bigg \{ - {i \over \hbar} \int dx
\xi(x)
\varpi(x) \bigg \} |\phi - {\xi \over 2} \rangle \langle \phi + {\xi
\over 2}|. }

Consider now the operator

\eqn \veinte{ \hat{I} := {1 \over 2} \int {\cal D} (2 \phi) |\phi \rangle
\langle - \phi|. }

Using \veinte\  we can define the field analog of the Grossmann
operator$^{22,14,15}$ as follows

$$ \hat{I}[\lambda,\mu] := \hat{\cal{U}}[\lambda,\mu] \hat{I} =
{1 \over 2} \int {\cal D}(2 \phi) exp \bigg \{ i \int dx \lambda(x) \bigg(
\phi(x) - {\hbar \mu(x) \over 2} \bigg) \bigg\} |\phi - \hbar
\mu \rangle \langle - \phi| $$

\eqn\vuno{ = {1 \over 2} \int {\cal D} (2 \xi) exp \bigg \{ i \int dx
\lambda(x) \xi (x) \bigg \} |\xi - {\hbar \mu \over 2} \rangle \langle
- \xi - {\hbar \mu \over 2}|. }

Comparing \dnueve\  with \vuno\  one quickly finds that

\eqn\vdos{ \hat{\Omega}[\phi,\varpi]= 2 \hat{I} \bigg[ {2 \varpi \over
\hbar}, - {2 \phi \over \hbar} \bigg] = \hat{\cal{U}}\bigg[ {2 \varpi
\over \hbar}, - {2 \phi \over \hbar}\bigg] 2 \hat{I}. }
Hence $\hat{\Omega}[0,0] = 2 \hat{I}$ and consequently

\eqn\vcuatro{ \hat{\Omega}[\phi,\varpi]= \hat{\cal{U}}\bigg[ {2\varpi
\over \hbar}, -{2 \phi \over \hbar} \bigg] \hat{\Omega}[0,0]. }

Simple manipulations show that \vcuatro\  can be also written in the
following form

$$\hat{\Omega}[\phi,\varpi] = \hat{\cal{U}} \bigg[ {\varpi \over
\hbar}, -{\phi \over \hbar} \bigg] \hat{\Omega}[0,0]
\hat{\cal{U}}^{^{^\dagger}}
\bigg[ {\varpi \over \hbar}, -{\phi \over \hbar} \bigg] $$   

\eqn\vcinco {= \hat{\cal{U}} \bigg[ {\varpi \over \hbar}, -{\phi \over
\hbar} \bigg] \hat{\Omega}[0,0] \hat{\cal{U}} \bigg[ -{\varpi \over  
\hbar}, {\phi \over \hbar} \bigg]. }

It is interesting to note that similarly as in the quantum mechanics$^{22}$
one can find the relation between the Stratonovich-Weyl quantizer
and the quantum field image of the Dirac $\delta$. Namely we have

$$ \hat{\delta} := W \bigg( \delta [\phi ] \delta \bigg[ {\varpi
\over 2 \pi\hbar} \bigg] \bigg) $$
\eqn\vseis{ = \int {\cal D} \phi {\cal D}({\varpi \over 2 \pi \hbar})
\delta [\phi]
\delta [{\varpi \over 2 \pi \hbar}]
\hat{\Omega}[\phi,\varpi] = \hat{\Omega}[0,0] = 2 \hat{I}.}

\vskip 1truecm
%%%%%%%%%%%%%%%%%%%%%%%%%%%%%
\subsec{ The Star Product}

Now we are in a position to define the Moyal $*$-product in field theory.
Let $F_1=F_1[\phi,\varpi]$ and $F_2=F_2[\phi,\varpi]$ be some functionals
on $\cal{Z}$ that correspond to the field operators $\hat{F}_1$ and
$\hat{F}_2$ respectively, i.e. $F_1[\phi,\varpi]=W^{-1}(\hat{F}_1)=Tr 
\big( \hat{\Omega} [\phi,\varpi] \hat{F}_1 \big)$ and
$F_2[\phi,\varpi]=W^{-1}(\hat{F}_2) = Tr \big(
\hat{\Omega}[\phi,\varpi] \hat{F}_2 \big)$. The question is to find the
functional which corresponds to the product $\hat{F}_1 \hat{F}_2$. This
functional will be denoted by $(F_1 \ast F_2)[\phi,\varpi]$ and we call
it the Moyal $\ast$-product$^{23,1}$. So we have

\eqn\vsiete { (F_1 \ast F_2)[\phi,\varpi]:= W^{-1}(\hat{F}_1 \hat{F}_2)=
Tr \bigg \{ \hat{\Omega}[\phi,\varpi] \hat{F}_1 \hat{F}_2 \bigg \}. }

Substituting \trece\  into \vsiete\  and performing simple calculations
one gets

$$ (F_1 \ast F_2) [\phi,\varpi] = \int {\cal D} (\phi ') {\cal D} (\phi'')
{\cal D} \big( {\varpi ' \over \pi \hbar} \big) {\cal D} \big( {\varpi
'' \over \pi \hbar}  \big) F_1[\phi', \varpi'] $$

\eqn\vocho{exp \bigg \{ {2i \over \hbar} \int dx \bigg((\phi - \phi
')(\varpi -
\varpi '') - (\phi - \phi '')(\varpi - \varpi ') \bigg) \bigg \} F_2[\phi
'',
\varpi ''].  }

To proceed further we introduce new variables $\varphi ' = \phi ' - \phi,
 \ \varphi '' = \phi '' - \phi, \ \upsilon ' = \varpi ' - \varpi, \upsilon
''=\varpi'' - \varpi .$ Using the expasion of $F_1[\phi ', \varpi '] = F_1[
\phi + \varphi ', \varpi + \upsilon ']$ and $F_2[\phi '', \varpi ''] = 
F_2[\phi + \varphi '', \varpi + \upsilon ''] $ in Taylor series and after
some laborious manipulations (in
principle very similar to the ones given in Ref. 21) we obtain

\eqn\treinta{ \big(F_1 *  F_2\big)[\phi,\varpi] =  F_1[\phi,\varpi]
exp\bigg\{{i\hbar\over 2} \buildrel{\leftrightarrow}\over {\cal P}\bigg\}
F_2[\phi,\varpi],}
where

\eqn\tuno{\buildrel{\leftrightarrow}\over {\cal P} := \int dx
\bigg({{\buildrel{\leftarrow}\over {\delta}}\over
\delta \phi(x)} {{\buildrel{\leftarrow}\over {\delta}}\over
\delta \varpi(x)} - {{\buildrel{\rightarrow}\over {\delta}}\over   
\delta \varpi(x)} {{\buildrel{\rightarrow}\over {\delta}}\over
\delta \phi(x)}\bigg)}
and

\eqn\tdos{
exp\bigg\{{i\hbar\over 2} \buildrel{\leftrightarrow} \over
{\cal P}\bigg\} = \sum_{l=0} {1 \over l!} \big( {i \hbar \over 2} \big)^l \int dx_1
\dots dx_l \omega^{i_1j_1} \dots \omega^{i_lj_l}
{{\buildrel{\leftarrow} \over {\delta^l}} \over {\delta
Z^{i_1}(x_1) \dots \delta Z^{i_l}(x_l)}} {{\buildrel{\rightarrow} \over
{\delta^l}} \over {\delta Z^{j_1}(x_1) \dots \delta Z^{j_l}(x_l)}},}
where $i_1, \dots, j_1, \dots = 1,2,$ $(Z^1,Z^2):=(\phi,\varpi),$
$(\omega^{ij}):=\pmatrix{0& 1\cr
-1&0\cr}$ and the overarrow indicates direction of action of the   
corresponding operator.

% \bigl( {0 \atop -1} {1\atop 0} \bigr)$

\vskip 1truecm
%%%%%%%%%%%%%%%%%%%%%%%%%%
\subsec{The Wigner Functional}

Finally we are going to define the Wigner functional. Let $\hat{\rho}$ be
the density operator of a quantum state. The functional
$\rho[\phi,\varpi]$ corresponding to $\hat{\rho}$ reads (see \docho)

\eqn\ttres{
\rho[\phi,\varpi]= Tr \bigg \{ \hat{\Omega}[\phi,\varpi]
\hat{\rho}\bigg \}
= \int {\cal D} \xi  exp \bigg\{ - { i \over \hbar} \int dx \xi(x)
\varpi(x)\bigg\} \langle \phi + {\xi \over 2} | \hat{\rho} | \phi - {\xi
\over 2} \rangle .}

Then in analogy to quantum mechanics$^{24,25,26}$ the
Wigner functional
$\rho_{_W}[\phi,\varpi]$ corresponding to the state $\hat{\rho}$ is
defined by

\eqn\tcuatro{\rho_{_W}[\phi,\varpi] : = \int {\cal D} ({\xi\over 2 \pi
\hbar}) exp \bigg\{ - { i \over \hbar} \int dx \xi(x)
\varpi(x)\bigg\} \langle \phi + {\xi \over 2} | \hat{\rho} | \phi - {\xi
\over 2} \rangle .}

For $\hat{\rho} = |\Psi\rangle \langle \Psi |$ \tcuatro\ gives

\eqn\tcinco{ \rho_{_W}[\phi,\varpi] = \int {\cal D} ({\xi \over 2 \pi
\hbar})
exp
\bigg\{ - {i \over \hbar} \int dx \xi(x) \varpi(x) \bigg\} \Psi^*[\phi -
{\xi \over 2}] \Psi[\phi + {\xi \over 2}]}
according to Ref. 12.

\noindent
[{\it Important Remark:} It must be noted that many, perhaps most, of 
our formulas are defined only formally. This resembles the case of the
path integral theory. We expect that further development of the formalism
presented in this paper will provide us with better mathematical frame.]

\vskip 2truecm   
%%%%%%%%%%%%%%%%%%%%%%%%%%%%%%%%%%%%%%%%%%%%%%%%%%%%%%%%%%%%%%%%%%%%%
%%%%%%%%%%%%%%%%%%%%%%%%%%%%%%%%%%%%%%%%%%%%%%%%%%%%%%%%%%%%%%%%%%%%%

\newsec{Free Scalar Field}

In this section we deal with free real scalar field of the action$^{27-30}$

\eqn\tseis {
{\cal S}[\phi] = \int dxdt {\cal L}(\phi(x,t), \partial_{\mu} \phi(x,t))
= - {1 \over 2} \int
dxdt(\partial^{\mu}\phi \partial_{\mu} \phi -m^2 \phi^2),}
where $\mu = 1, \dots, 4$ (we put $c=1$).
The conjugate field momentum is $\varpi (x,t) = {\partial {\cal L} \over \partial(
\partial \dot{\phi}) } = \dot{\phi}(x,t),$  where $\dot{\phi} \equiv {\partial
\phi \over \partial t}.$ Then the Hamiltonian reads

\eqn\tocho { H[\phi, \varpi] = {1 \over 2} \int dx \bigg( \varpi (x,t)^2 +
(\nabla \phi(x,t))^2 + m^2 \phi (x,t)^2 \bigg). }

The field $\phi (x,t)$ satisfies the Klein-Gordon equation
$(\partial_{\mu} \partial^{\mu} - m^2) \phi (x,t) = 0.$
For $\phi (x,t)$ and $\varpi (x,t)$ we have the usual Poisson brackets

$$\{ \phi (x,t), \varpi (y,t) \} = \delta (x-y), $$

\eqn\cuarenta { \{ \phi (x,t), \phi (y,t) \} = 0 = \{ \varpi (x,t), \varpi
(y,t) \}. }

Consider the standard expansion of $\phi$ and $\varpi$

$$\phi(x,t) = {1 \over (2 \pi)^{3/2}} \int dk \bigg( {\hbar
\over 2 \omega(k)} \bigg)^{1/2} \bigg( a(k,t) exp\{ ikx \}
+ a^{*}(k,t) exp \{ -ikx \} \bigg), $$

\eqn\cuno{\varpi (x,t) = \dot{\phi} (x,t) = {1 \over (2 \pi)^{3/2}} \int
dk \ i \bigg( {\omega (k) \hbar \over 2 } \bigg)^{1/2} \bigg(
a^{*}(k,t) exp\{ -ikx \} - a(k,t) exp \{ ikx \} \bigg),  }
where $\omega (k) = \sqrt{k^2 + m^2},$ $a(k,t) = a(k)exp \big\{ -i
\omega (k)t \big\}$ and $kx \equiv  k_j x_j,$ $j=1,2,3.$

From \cuno\  we get

\eqn\cdos { a(k,t) = {1 \over (2 \pi)^{3/2}(2 \omega (k)\hbar)^{1/2} }
\int dx \  exp \big\{ -ikx \big\} \bigg(\omega (k) \phi (x,t) + i \varpi
(x,t)\bigg).}

One can easily check that  \cuarenta\  and \cdos\  give

$$ \{ a(k,t), a^{*}(k',t) \} = -{i \over \hbar} \delta (k - k'), $$

\eqn\ctres { \{ a(k,t), a(k',t) \} = 0= \{ a^{*}(k,t), a^{*}(k',t) \}.  }

\vskip 1truecm
%%%%%%%%%%%%%%%%%%%%%%%%%
\noindent
\subsec{Canonical Transformation}

Then we introduce new canonical field variables$^{31,32}$

\eqn\ccuatro{
Q(k,t):= \big( {\hbar \over 2 \omega (k)}
\big)^{1/2}\bigg(a(k,t) + a^{*}(k,t) \bigg), \ \
P(k,t):= i \big( {\hbar \omega (k) \over 2}
\big)^{1/2}\bigg(a^{*}(k,t) - a(k,t)\bigg). }
Hence

\eqn\ccinco { a(k,t)= \big( {\omega (k) \over 2 \hbar} \big)^{1/2} \bigg(
Q(k,t) + {i \over \omega (k)} P(k,t) \bigg). }

From \ctres\  and \ccuatro\  one has
$$ \{ Q(k,t),P(k',t) \} = \delta (k - k'), $$

\eqn\cseis { \{ Q(k,t),Q(k',t) \} = 0 = \{ P(k,t),P(k',t) \}. }

Inserting \cdos\ into \ccuatro\  we obtain

$$ Q(k,t) = {1 \over (2 \pi)^{3/2} \omega (k)} \int dx \bigg(\varpi
(x,t)sin(kx) + \omega (k) \phi (x,t) cos(kx)\bigg),$$

\eqn\csiete { P(k,t) = {1 \over (2 \pi)^{3/2}} \int dx \bigg( \varpi
(x,t)cos(kx) - \omega (k) \phi (x,t) sin(kx) \bigg). }

Then the inverse transformation reads

$$ \phi(x,t) = {1 \over (2 \pi)^{3/2}} \int dk \bigg(Q(k,t)
cos(kx) - {P(k,t) \over \omega (k)} sin(kx) \bigg),$$

\eqn\cocho { \varpi(x,t) = {1 \over (2 \pi)^{3/2}} \int dk \bigg(\omega(k)
Q(k,t) sin(kx) + P(k,t) cos(kx)\bigg). }

We now consider the field at the instant $t=0$.
From \cseis\ it follows that \csiete\ defines a linear canonical
transformation. Consequently the measure ${\cal D} \phi {\cal D} \varpi =
{\cal D}Q {\cal D}P$ and all the formalism given before can be easily
expressed in terms of new variables $Q$ and $P$.

\vskip 1truecm
%%%%%%%%%%%%%%%%%%%%%%%%%%%%%%%%
\noindent
\subsec{The Stratonovich-Weyl Quantizer}

Eqs. \trece\ and \catorce\ can be rewritten in the following form

\eqn\cnueve{ \hat{F} = \int {\cal D} Q {\cal D} ({P\over 2 \pi \hbar})
F[\phi[Q,P], \varpi[Q,P]] \hat{\Omega}[\phi[Q,P], \varpi[Q,P]]}
and

$$\hat{\Omega} [\phi[Q,P], \varpi[Q,P]] = \int
{\cal D} ({\hbar \lambda \over 2 \pi}){\cal D} \mu exp \bigg \{ -i \int dk
\bigg(\lambda(k) Q(k) + \mu (k) P(k) \bigg) \bigg \} $$

\eqn\cincuenta{ {\rm exp}\bigg\{i\int dk \bigg(
\lambda({k}) \hat{Q}({k}) +
\mu({k}) \hat{P}({k})\bigg)\bigg\} }
with $\hat{Q}$ and $\hat{P}$ the field operators and corresponding states $|Q \rangle$
and $|P \rangle$ satisfying the
relations: $\hat{Q}(k) | Q \rangle = Q(k) |Q\rangle ,$
$\hat{P}(k) | P \rangle = P(k) |P\rangle,$
$\int {\cal D} Q |Q \rangle \langle Q| = \hat{1}$ and
$\int {\cal D} ({P \over 2 \pi \hbar}) |P \rangle \langle P| = \hat{1}.$

Further on we denote the Stratonovich-Weyl quantizer
$\hat{\Omega}[\phi[Q,P],\varpi[Q,P]]$ simply by $\hat{\Omega}[Q,P]$.
Then

$$ \hat{\Omega}[Q,P] = \int {\cal D} ({\eta \over 2 \pi
\hbar}) exp \bigg \{ -{i \over \hbar} \int dk \eta(k) Q(k) \bigg \}
|P + {\eta \over 2} \rangle \langle P - {\eta \over 2}| $$

\eqn \cidos { = \int {\cal D} \xi exp \bigg \{ - {i \over \hbar} \int dk
\xi(k) P(k) \bigg \} |Q - {\xi \over 2} \rangle \langle Q + {\xi
\over 2}|. }

It is evident how to write the Grossmann operator within the
$(Q,P)$ formalism, so we do not consider this here.

\vskip 1truecm
%%%%%%%%%%%%%%%%%%%%%%%%%%%%%%%%%%%%%%
\noindent
\subsec{Star Product and Wigner Functional}

One easily shows that the Moyal $\ast$-product can be now expressed by  

$$ \big(F_1 *  F_2\big)[Q,P] = F_1[Q,P] exp\bigg\{{i\hbar\over 2}
\buildrel{\leftrightarrow} \over {\cal P}\bigg\} F_2[Q,P], $$

\eqn\cicuatro{\buildrel{\leftrightarrow}\over {\cal P} := \int dk
\bigg({{\buildrel{\leftarrow}\over {\delta}}\over \delta Q(k)}
{{\buildrel{\rightarrow}\over {\delta}}\over \delta P(k)}
- {{\buildrel{\leftarrow}\over {\delta}}\over \delta P(k)}
{{\buildrel{\rightarrow}\over {\delta}}\over \delta Q(k)}\bigg).}

Finally, the Wigner functional in the $(Q,P)$ formalism is given by

\eqn\cicinco{ \rho_{_W}[Q,P] = \int {\cal D} ({\xi\over 2 \pi \hbar})  exp
\bigg \{ - { i \over \hbar} \int dk \xi(k)
P(k)\bigg\} \langle Q + {\xi \over 2} | \hat{\rho} | Q - {\xi
\over 2} \rangle, }
where $\rho_{_W}[Q,P]$ means $\rho_{_W}[\phi [Q,P], \varpi [Q,P]]$.
Then for $\hat{\rho} = |\Psi\rangle\langle \Psi |$ one has

\eqn\ciseis{ \rho_{_W}[Q,P] = \int {\cal D} ({\xi \over 2 \pi
\hbar}) exp \bigg\{ - {i \over \hbar} \int dk \xi(k) P(k) \bigg\} \Psi^*[Q -
{\xi \over 2}] \Psi[Q + {\xi \over 2}]. }

Now we are going to find the Wigner functional for the ground state $| \Psi_0
\rangle$. From the very definition of $|\Psi_0\rangle$

\eqn\cisiete { \hat{a}(k)|\Psi_0 \rangle = 0.  }

Substituting \ccinco\ and using the $Q$ representation we get the functional
equation

\eqn\ciocho { \bigg( {Q}(k) + {\hbar \over \omega (k)} {\delta \over
\delta Q(k)} \bigg) \Psi_0[Q] = 0.}
Hence

\eqn\cinueve { \Psi_0[Q] \propto exp \bigg\{ -{1 \over 2 \hbar} \int dk   
\omega (k) Q^{2}(k) \bigg\}. }

Finally, inserting \cinueve\ into \ciseis\ and performing some
straightforward calculations one finds the Wigner functional
${\rho_{_{W0}}}$ of the ground state to be

\eqn\sesenta { \rho_{_{W0}}[Q,P] \propto exp \bigg\{ -{1 \over \hbar}
\int
{dk \over \omega (k)} \bigg( P^{2}(k) + \omega ^2 (k) Q^2 (k) \bigg)
\bigg \}.}

Employing \csiete\ we obtain the Wigner functional of the ground state in
terms of $(\phi, \varpi)$

\eqn\suno{ \rho_{_{W0}} [\phi,\varpi] \propto
exp \bigg \{ - {1 \over \hbar} \int dx \bigg( \phi (x) (\sqrt{ -
\nabla^2_x + m^2}) \phi(x) \bigg) + \bigg( \varpi(x) ( \sqrt{ -
\nabla^2_x + m^2})^{-1} \varpi(x) \bigg) \bigg \} }
according to Ref. 12.

Comparing the Wigner functional (3.21) with the harmonic oscillator Wigner
function given in Refs. 26, 33 and 34 we conclude
that the former one represents
the Wigner function of infinite number of harmonic oscillators. It's
nothing strange as the variables $Q$ and $P$ are the field theoretical
analogs of normal coordinates and their conjugate momenta.

\vskip 1truecm
%%%%%%%%%%%%%%%%%%%%%%%%%%%%%%
\noindent
\subsec{Oscillator Variables and Ordering}

Now one can easily find the Wigner functionals corresponding to higher
states. Let $ | \dots k' \dots k'' \dots k^{(n)} \dots \rangle = \dots
\hat{a}^{^{\dagger}}(k') \dots \hat{a}^{^{\dagger}} (k'')
\dots \hat{a}^{^{\dagger}}(k^{(n)}) \dots  |\Psi_0 \rangle $ be the higher
quantum
state of the scalar field. The density operator $\hat{\rho}$ reads

\eqn\sdos{ \hat{\rho}_{ (\dots k' \dots k'' \dots k^{(n)} \dots )} \propto \dots
\hat{a}^{^{\dagger}}(k') \dots \hat{a}^{^{\dagger}} (k'')
\dots \hat{a}^{^{\dagger}}(k^{(n)}) \dots  |\Psi_0 \rangle
\langle \Psi_0|  \dots \hat{a}(k^{(n)}) \dots \hat{a}(k'') \dots
\hat{a}(k') \dots }
with $\hat{a}(k)$ and $\hat {a}^{^{\dagger}}(k)$ being the annihilation
and creation operators, respectively.
Hence the corresponding  Wigner functional is

$$
{\rho}_{_{W( \dots k' \dots k'' \dots k^{(n)}\dots )}}[a,a^*] \propto \dots
*a^*(k') * \dots * a^*(k'') * \dots * a^*(k^n) * \dots *
\rho_{_{W0}}[a,a^*]
* \dots * a(k^n)* $$

\eqn\stres{\dots * a(k'')* \dots * a(k') * \dots , }
where by (3.7) and (3.21)

\eqn\scuatro{ \rho_{_{W0}} [a,a^*] \propto exp \bigg \{ -2 \int dk a^*(k)
a(k) \bigg \}. }

The Moyal $\ast$-product operator in terms of $a(k)$ and $a^*(k)$ can be
written as follows$^7$

\eqn\scinco{ *   = exp\bigg\{{i\hbar\over 2}
\buildrel{\leftrightarrow} \over {\cal P}\bigg\} = exp \bigg\{  
{1 \over 2}\int dk 
\bigg({{\buildrel{\leftarrow}\over {\delta}}\over \delta a(k)}
{{\buildrel{\rightarrow}\over {\delta}}\over \delta a^*(k)}
- {{\buildrel{\leftarrow}\over {\delta}}\over \delta a^*(k)}
{{\buildrel{\rightarrow}\over {\delta}}\over \delta a(k)}\bigg) \bigg\}.}
  
Consequently, any $*$-product of $a^*$'s or $a$'s can be rewritten as usual
product of functions and the Wigner functional \stres\ is given by
(compare with Ref. 33)

$$ {\rho}_{_{W( \dots k' \dots k'' \dots k^{(n)} \dots)}}[a,a^*]$$

\eqn\sseis{ \propto \dots a^*(k')  \dots  a^*(k'')  \dots  a^*(k^{(n)})  \dots
* \rho_{_{W0}}[a,a^*] * \dots a(k^{(n)}) \dots  a(k'') \dots  a(k')  \dots
 \ \ .}

An interesting question arises if we are able to define the normal
ordering within deformation quantization formalism for field theory.
Indeed it can be easily done by a suitable generalization of the results
by Agarwal and Wolf$^{35}$ (see also Ref. 26). Let $F[Q,P]$ be a functional
over ${\cal Z}$. Define the functional $ F_{\cal N}[Q,P] $ as follows

$$ F_{\cal N}[Q,P] := \hat{\cal N} F[Q,P],
$$
where 

\eqn\ssiete{
\hat{\cal N} := exp \bigg\{ - {\hbar \over 4} \int dk \bigg( \omega(k) {
\delta^2 \over \delta P^2(k)} + {1 \over \omega(k)} { \delta^2 \over
\delta Q^2(k)} \bigg) \bigg\} 
 = exp \bigg\{ - {1\over 2} \int dk { \delta^2 \over \delta
a(k) \delta a^*(k)} \bigg \}. }
This formula was first obtained by Dito$^7$.

Then the Weyl image of $F_{\cal N} [Q,P]$ gives the normal ordering of the
Weyl image of $F[Q,P]$

\eqn\socho{ : W[ F[Q,P]] : \ \buildrel{df}\over {=} W [F_{\cal N}[Q,P]]
\buildrel{df}\over {=} W_{\cal N} (F[Q,P])}

It is worthwhile to note some interesting property of normal ordering

\eqn\secuatro{ W_{\cal N}^{-1} \bigg( :(:\hat{F}_1:)(:\hat{F}_2:): \bigg)
= W^{-1} \bigg( (: \hat {F}_1 :)(: \hat{F}_2 :) \bigg)  = \bigg(\hat {\cal
N} W^{-1} (\hat {F}_1) \bigg) * \bigg( \hat {\cal N} W^{-1} (\hat {F}_2)   
\bigg). }

\vskip 2truecm
%%%%%%%%%%%%%%%%%%%%%%%%%%%%%%%%

\noindent
{\it Example: The Hamiltonian}
  
From (3.2) and (3.4) we have

\eqn\snueve{ H[a,a^*] = \int dk \hbar \omega(k) a^*(k)a(k). }
Then 

\eqn\setenta{ H_{\cal N}[a,a^*] =
\int dk \hbar \omega(k) a^*(k)a(k)
- {1 \over 2} \int dk \hbar \omega(k) \delta(0),  }
and by the quantum version of Eq. (3.6) we get

$$ : \hat{H} : \ = W \bigg( H_{\cal N}[a,a^*] \bigg)=
{1 \over 2} \int dk \hbar \omega(k) \bigg(\hat{a}^{^{\dagger}}(k)
\hat{a}(k) + \hat{a}(k) \hat{a}^{^{\dagger}}(k) \bigg)
- {1 \over 2} \int dk \hbar \omega(k) \delta(0)$$

$$
= \int dk \hbar \omega(k) \hat{a}^{^{\dagger}}(k)
\hat{a}(k)
+ {1 \over 2} \int dk \hbar \omega(k) \delta(0)
- {1 \over 2} \int dk \hbar \omega(k) \delta(0) $$

\eqn\seuno{ = \int dk \hbar \omega(k) \hat{a}^{^{\dagger}}(k)
\hat{a}(k), }
where $\delta(0)$ means here the Dirac delta in three dimensions.

The eigenvalue Schr\"odinger equation reads

\eqn\sedos{ H_{\cal N} * \rho_{_{W}} = E \rho_{_{W}}. }

One immediately finds that the vacuum energy is zero i.e.,

\eqn\setres{ H_{\cal N} * \rho_{_{W0}} = 0}   
as it should be.

\vskip 1truecm
%%%%%%%%%%%%%%%%%%%%%%%%%%%%%%%%

\noindent
\subsec{Time Evolution}

Finally we consider the time evolution of Wigner functional.
The von Neumann-Liouville equation reads

\eqn\secinco{ {\partial \rho_{_W}[a,a^*;t] \over \partial t} =
\{\hat{\cal N}H[a,a^*], \rho_{_W}[a,a^*;t] \}_{\cal M}, }
where $\{\cdot,\cdot\}_{\cal M}$ stands for the Moyal bracket

\eqn\seseis{ \{F_1,F_2\}_{\cal M} = {1 \over i \hbar} \big(F_1 * F_2 - F_2
* F_1 \big) = F_1 {2 \over \hbar}  \sin \big({ \hbar\over 2}
\buildrel{\leftrightarrow}\over {\cal P} \big) F_2. }

For the Hamiltonian given by (3.31), the Moyal bracket
in \secinco\ reduces to the Poisson bracket. So we have   

\eqn\sesiete{ {\partial \rho_{_W}[a,a^*;t] \over \partial t} =
\{\hat{\cal N}H[a,a^*], \rho_{_W}[a,a^*;t] \}. }

This result within the $(\phi,\varpi)$-formalism has been found previously
by Curtright and Zachos$^{12}$.

\vskip 2truecm
%%%%%%%%%%%%%%%%%%%%%%%%%%%%%%%%%%%%%%%%%%%%%%%%%%%%%%%%%%%%%%%%%%%%%
%%%%%%%%%%%%%%%%%%%%%%%%%%%%%%%%%%%%%%%%%%%%%%%%%%%%%%%%%%%%%%%%%%%%%

\newsec{Free Electromagnetic Field}

In this section we consider deformation quantization for free
electromagnetic field. This case seems to be very interesting as it
is the simplest example of a field theory with constraints. (For details
see Refs. 27-31).

We choose the temporal gauge where the fourth (temporal) component of the 
gauge potential $A_4=0.$ The canonical components are the potential
$A=(A_1,A_2,A_3)$ and its conjugate momentum $\varpi = \dot{A} = -E=
-(E_1,E_2,E_3)$ with $E$ being the electric field$^{27-31}$.

Fields $A$ and $E$ satisfy usual relations

$$\{A_i(x,t), E_j(y,t) \} = - \delta_{ij} \delta(x - y), $$

\eqn\seocho{ \{A_i(x,t), A_j(y,t) \} = 0 = \{E_i(x,t),E_j(y,t)\}, \ \ i,j= 
1,2,3. }

The standard expansion of the field variables at $t=0$ reads

$$
A_j(x) = { 1\over (2 \pi)^{3/2}} \int dk \bigg({ \hbar
\over 2 \omega(k)}\bigg)^{1/2} \bigg( a_j(k) exp\big(ikx \big) +
a^*_j(k) exp \big(-ikx \big) \bigg),  $$

\eqn\senueve{ 
E_j(x) = { 1\over (2 \pi)^{3/2}} \int dk \ i
\bigg({
\hbar \omega(k)
\over 2}\bigg)^{1/2} \bigg( a_j(k) exp\big(ikx \big) -
a^*_j(k) exp \big(-ikx \big) \bigg), }
where $\omega(k) = |k|.$

From Eq.  \senueve\ one gets  

\eqn\ochenta{ a_j (k) = {1 \over (2 \pi)^{3/2} ( 2 \hbar
\omega(k))^{1/2}} \int dx \ exp \big\{ -ikx  \big\} \bigg( \omega(k)
A_j(x)
- i E_j(x) \bigg). }
Then by Eqs. \seocho\ and \senueve\ we have 

$$\{a_i(k), a^*_j(k') \} = - {i \over \hbar} \delta_{ij} \delta(k -
k'), $$

\eqn\ouno{ \{a_i(k), a_j(k') \} = 0 = \{a^*_i(k),a^*_j(k')\}.}

Similarly as in the previous section one can introduce new
coordinates $Q$ and their conjugate $P$ momenta$^{31,32}$

$$Q_j(k) := \big({ \hbar \over 2 \omega(k)}\big)^{1/2} \bigg( a^*_j(k)
+a_j(k) \bigg),$$

$$ P_j(k) := i\big({ \hbar \omega(k) \over 2}\big)^{1/2} \bigg(
a^*_j(k) - a_j(k) \bigg),$$

$$ \{ Q_i(k), P_j(k') \} = \delta_{ij} \delta(k-k'), $$

\eqn\odos{ \{ Q_i(k), Q_j(k') \} = 0 = \{ P_i(k),P_j(k') \}.}

Comparing \seocho\ with \odos\ we can observe that the transformation between
coordiantes ($A$,$-E$) and $(Q,P)$ is a canonical one.

Then one gets 
$$ 
Q_j(k) = { 1\over (2 \pi)^{3/2}\omega(k)} \int dx   \bigg(
 \omega(k)A_j(x) cos(kx) - E_j(x) sin(kx) \bigg),
$$

\eqn\otres{P_j(k) = -{ 1\over (2 \pi)^{3/2}} \int dx   \bigg(
 \omega(k)A_j(x) sin(kx) + E_j(x) cos(kx) \bigg)}
and

$$ 
A_j(x) = { 1\over (2 \pi)^{3/2}} \int dk \  \bigg(
 Q_j(k) cos(kx) - {P_j(k) \over \omega(k)} sin(kx) \bigg),
$$

\eqn\ocuatro{E_j(x) = -{ 1\over (2 \pi)^{3/2}} \int dk  \bigg(
 \omega(k)Q_j(k) sin(kx) + P_j(k) cos(kx) \bigg).}

In the standard way we can now split the field objects into their transverse
$(T)$ and longitudinal $(L)$ components.
To this end we write $a_j$ in the following form

\eqn\ocinco { a_j(k)=h_{jl}(k)a_l(k) + {k_jk_l \over |k|^2}a_l(k), }
where $h_{jl}(k)$ is the projector on the space perpendicular to $k$ i.e.,
$h_{jl} = \delta_{jl} - {k_jk_l \over |k|^2} $. Introducing two polarization
vectors $e^1(k)$ and $e^2(k)$ such that 

\eqn\oseis{ e^{\alpha}_i(k) e^{\alpha '}_i(k) = \delta_{\alpha \alpha
'}, \ \ \ k_i e^{\alpha}_i(k) = 0, \ \ \ \alpha, \alpha '=1,2 }
one can write 

\eqn\osiete{ e^{\alpha}_i(k) a_{T \alpha}(k):= h_{ij}(k)a_j(k)
\Rightarrow a_{T \alpha}(k) = e^{\alpha}_j(k) a_j(k)  }

Defining also $a_L (k):= {k_j \over |k|} a_j (k)$ we have 

\eqn\oocho{ a_i (k) = e^{\alpha}_i(k)a_{T \alpha}(k) + {k_i \over |k|}
a_L(k). }

One quickly shows that

\eqn\onueve{ \{ a_{T \alpha}(k),a^*_{T \alpha '}(k') \} = -{i \over
\hbar} \delta_{\alpha \alpha '} \delta(k-k'), \ \ \
\{a_L(k),a^*_L(k')\} = -{i \over \hbar} \delta(k-k'), }
with all the remaining Poisson brackets being zero.  Substituting \oocho\
and \onueve\ into (4.5) we obtain the expressions for the $T$ and
$L$-components of $Q$ and $P$. Then inserting \oocho\ into (4.2) one gets   

$$
A_j(x) = { 1\over (2 \pi)^{3/2}} \int dk \  \bigg({ \hbar \over 2 
\omega(k)} \bigg)^{1/2} \bigg\{ e^{\alpha}_j(k) \bigg( a_{T \alpha}(k)
exp\big(ikx \big) + a^*_{T \alpha}(k) exp \big(-ikx \big) \bigg)
$$

$$+ {k_j \over
|k|} \bigg(a_L(k) - a^*_L(-k) \bigg) exp \big(ikx \big) \bigg\}, 
$$

$$
E_j(x) = { 1\over (2 \pi)^{3/2}} \int dk \ i \bigg({ \hbar \omega(k)
\over 2}\bigg)^{1/2} \bigg\{ e^{\alpha}_j(k)[ a_{T \alpha}(k)
exp\big(ikx
\big) - a^*_{T \alpha}(k) exp \big(-ikx \big) \bigg)
$$
\eqn\noventa{ + {k_j \over |k|}
\bigg(a_L(k) + a^*_L(-k) \bigg) exp \big(ikx \big) \bigg \}. }

Then the Hamiltonian of the electromagnetic field reads

$$H = {1 \over 2} \int dx \bigg( E^2(x) + (\nabla \times
A(x))^2 \bigg) $$

$$ = \int dk \hbar \omega(k) a^*_{T \alpha}(k)a_{T \alpha}(k) +
{1 \over 4} \int dk \hbar \omega(k) (a_L(k)+a^*_L(-k))^*(a_L(k)+a^*_L(-k))
$$

\eqn\nuno{ = {1 \over 2} \int dk \big( P_T^2(k) + \omega^2(k) Q_T^2(k) \big) 
+ { 1\over 8} \int dk \big( (P_L(k)-P_L(-k))^2 + \omega^2(k) (Q_L(k)+
Q_L(-k))^2 \big). }

Given $H$ we can solve the Hamiltonian equations for $Q_j(k,t)$ and $P_j(k,t)$.
Simple calculations show that for the transversal part we obtain

$$
Q_{T \alpha}(k,t) = Q_{T \alpha}(k) \cos (\omega(k)t) + {P_{T \alpha}(k) \over \omega(k)} \sin 
(\omega(k)t),
$$

\eqn\nunoaux{
P_{T \alpha}(k,t) = - \omega(k) Q_{T \alpha}(k) \sin (\omega(k)t) + P_{T \alpha}(k) \cos  
(\omega(k)t).}
While the solutions for the longitudinal part are

$$
Q_L(k,t)= Q_L(k) + {1 \over 2} \bigg( P_L(k) - P_L(-k)\bigg) t
$$

\eqn\nunoauxu{
P_L(k,t) = P_L(k) - {1 \over 2} \omega^2(k) \bigg( Q_L(k) + Q_L(-k) \bigg) t.}

Consequently time evolution of $A_j$ and $E_j$ is given by

$$
A_j(x,t) = {1 \over (2 \pi)^{3/2}} \int dk \ \bigg\{ \bigg( Q_{T \alpha}(k,t) \cos (kx)
- { P_{T \alpha}(k,t) \over \omega(k)} \sin (kx) \bigg) e^{\alpha}_j(k) 
$$

$$ + {k_j \over \omega(k)}\bigg(  \big(Q_L(k) + P_L(k)t \big) \cos (kx)
+ \big( \omega(k) Q_L(k)t - {P_L(k)\over \omega(k)} \big) \sin (kx) \bigg) \bigg\}
$$ 

$$
E_j(x,t)= - \partial_t A_j(x,t) = - {1 \over (2 \pi)^{3/2}} \int dk \ \bigg\{ \omega(k)
Q_{T \alpha}(k,t) \sin (kx)
+ P_{T \alpha}(k,t) \cos (kx) \bigg) e^{\alpha}_j(k) 
$$

\eqn\solut{
 - {1 \over (2 \pi)^{3/2}} \int dk {k_j \over \omega(k)}
\bigg( P_L(k) \cos (kx) + \omega(k) Q_L(k) \sin (kx) \bigg\}.}

Now the constraint (the Gauss equation)

\eqn\ndos { \partial_jE_j(x)=0, } 
is equivalent to the following constraint

\eqn\ntres{ a_L(k) + a^*_L(-k)=0 \Longleftrightarrow  
 Q_L(k) + Q_L(-k)  + i \big({P_L(k) - P_L(-k) \over \omega(k)} \big)= 0,}
which is equivalent to the conditions

\eqn\constra{
Q_L(k) + Q_L(-k) = 0, \ \ \ \ \ P_L(k) - P_L(-k) = 0.}

Note that
the gauge transformation $A_j(x,t) \to A_j(x,t) + \partial_j \Lambda(x)$
produces the additional longitudinal term in $A_j$ of the form $\int dk
k_j i \bigg(b(k) + b^*(-k)\bigg) exp\{ikx\}$ which of course doesn't
change
$E_j(x,t)$,
$\nabla \times A(x,t)$, $H$, etc.

\vskip 1truecm
%%%%%%%%%%%%%%%%%%%%%%%%%%%%%%%%

\subsec{The Stratonovich-Weyl Quantizer for the Electromagnetic Field}

Consider now the Weyl quantization of electromagnetic field.  To this end
we deal with the fields at the moment $t=0$. From the previous section one
can conclude that it is very useful to employ $(P,Q)$ or $a$-formalisms. 
Let $F=F[Q,P]$ be a functional on electromagnetic field phase space ${\cal
Z}_E$. Then according to the Weyl rule we assigne to the functional $F$
the following operator $\hat{F}$

\eqn\ncuatro{
\hat{F} = W(F[Q,P]):= \int {\cal D} Q {\cal D} ({P\over 2 \pi \hbar})
F[Q,P] \hat{\Omega}[Q,P]}
with the Stratonovich-Weyl quantizer defined similarly
as before

$$
\hat{\Omega} [Q,P] = \int 
{\cal D} ({\hbar \lambda \over 2 \pi}){\cal D} \mu exp \bigg \{ -i \int dk
\bigg(\lambda(k) Q(k) + \mu (k) P(k) \bigg) \bigg \} $$

\eqn\ncinco{ {\rm exp}\bigg\{i\int dk \bigg(\lambda({k}) \hat{Q}({k}) +
\mu({k}) \hat{P}({k})\bigg)\bigg\}, }
where $\lambda(k)Q(k) := \lambda_{T \alpha}(k) Q_{T \alpha}(k) + \lambda_L(k) Q_L(k)$,
etc. and  all measures used in the integrals contain the longitudinal (${\cal
D}Q_L$) and the transverse (${\cal D}Q_T$) components and $\hat{Q}(k)$ and $\hat{P}(k)$ 
are field operators 

$$ \hat{Q}(k) |Q \rangle = Q(k) |Q \rangle, \ \ \ \ 
\hat{P}(k) |P \rangle = {P}(k) |P \rangle, $$

$$ \hat{Q}(k) = \big(\hat{Q}_T(k),\hat{Q}_L(k) \big), \ \  \hat{P}(k) = \big(\hat{P}_T(k),
\hat{P}_L(k) \big), \  |Q\rangle = |Q_T\rangle \otimes |Q_L\rangle, \  |P\rangle =  |P_T\rangle\
\otimes |P_L\rangle, $$
 
\eqn\nseis{
\int {\cal D} Q |Q \rangle \langle Q| = \hat{1} \ \ {\rm and} \
\  \int {\cal D} ({P \over 2 \pi \hbar}) |P \rangle \langle P| = \hat{1}. 
}

The commutation relations for $\hat{Q}$ and $\hat{P}$ operators read

\eqn\nsiete{ [\hat{Q}_{T \alpha}(k), \hat{P}_{T \alpha '}(k')] = i \hbar 
\delta_{\alpha \alpha '} \delta(k - k'), \ \ \ 
[\hat{Q}_{L}(k), \hat{P}_{L}(k')] = i \hbar  \delta(k - k')}
and all remaining commutators are zero.

Then by (4.5) the relation between $\hat{Q}$ and $\hat{P}$ operators and the 
annihilation and creation operators $\hat{a}$ and $\hat{a}^{^{\dagger}}$ is

$$\hat{Q}_{T,L}(k) := \big({ \hbar \over 2 \omega(k)}\big)^{1/2} \bigg( \hat{a}^{^{\dagger}}_{T,L}
(k) + \hat{a}_{T,L}(k) \bigg),$$

\eqn\nocho{ \hat{P}_{T,L}(k) := i\big({ \hbar \omega(k) \over 2}\big)^{1/2} \bigg(
\hat{a}^{^{\dagger}}_{T,L}(k) - \hat{a}_{T,L}(k) \bigg).}
We have the usual commutation relations 
 
\eqn\nnueve{
[\hat{a}_{T\alpha}(k), \hat{a}^{^{\dagger}}_{T \alpha '}(k')] =  \delta_{\alpha \alpha ' }
\delta(k - k'), \ \   \ \ \ 
[\hat{a}_{L}(k), \hat{a}^{^{\dagger}}_{L}(k')] =  
\delta(k - k') }
with all remaining commutators being zero.

The Stratonovich-Weyl quantizer (4.22) can be rewritten as before in the
form of (3.14) and it has the standard properties analogous to (2.10),
(2.11) and (2.12).

\vskip 1truecm
%%%%%%%%%%%%%%%%%%%%%%%%%%%

\subsec{ The Star Product}

From (4.21)  we get 

\eqn\cien{ F[Q,P] = W^{-1}(\hat{F}) = Tr \bigg\{ \hat{\Omega}[Q,P] \hat{F}
\bigg\}.}

The Moyal $*$-product in the case of electromagnetic field theory can be
defined in a similar way as for the scalar field. Let $F_1[Q,P]$ and
$F_2[Q,P]$ the functionals on ${\cal Z}_E$ and let $\hat{F}_1$ and
$\hat{F}_2$ be their corresponding operators.  Then the analogous
calculations as in Sec. 2 lead to the result

$$ \big(F_1 *  F_2\big)[Q,P] = F_1[Q,P] exp\bigg({i\hbar\over 2}
\buildrel{\leftrightarrow} \over {\cal P}\bigg) F_2[Q,P], $$

$$\buildrel{\leftrightarrow}\over {\cal P} := \int dk 
\bigg({{\buildrel{\leftarrow}\over {\delta}}\over \delta Q(k)} 
{{\buildrel{\rightarrow}\over {\delta}}\over \delta P(k)}
- {{\buildrel{\leftarrow}\over {\delta}}\over \delta P(k)} 
{{\buildrel{\rightarrow}\over {\delta}}\over \delta Q(k)}\bigg)$$

$$ = \int dk 
\bigg({{\buildrel{\leftarrow}\over {\delta}}\over \delta Q_T(k)} 
{{\buildrel{\rightarrow}\over {\delta}}\over \delta P_T(k)}
- {{\buildrel{\leftarrow}\over {\delta}}\over \delta P_T(k)} 
{{\buildrel{\rightarrow}\over {\delta}}\over \delta Q_T(k)}\bigg)
+ \int dk 
\bigg({{\buildrel{\leftarrow}\over {\delta}}\over \delta Q_L(k)} 
{{\buildrel{\rightarrow}\over {\delta}}\over \delta P_L(k)}
- {{\buildrel{\leftarrow}\over {\delta}}\over \delta P_L(k)} 
{{\buildrel{\rightarrow}\over {\delta}}\over \delta Q_L(k)}\bigg)
$$

\eqn\cienuno{
=- {i \over \hbar} \bigg\{ \int dk 
\bigg({{\buildrel{\leftarrow}\over {\delta}}\over \delta a_T(k)} 
{{\buildrel{\rightarrow}\over {\delta}}\over \delta a^*_T(k)}
- {{\buildrel{\leftarrow}\over {\delta}}\over \delta a^*_T(k)} 
{{\buildrel{\rightarrow}\over {\delta}}\over \delta a_T(k)}\bigg)
+ \int dk 
\bigg({{\buildrel{\leftarrow}\over {\delta}}\over \delta a_L(k)} 
{{\buildrel{\rightarrow}\over {\delta}}\over \delta a^*_L(k)}
- {{\buildrel{\leftarrow}\over {\delta}}\over \delta a^*_L(k)} 
{{\buildrel{\rightarrow}\over {\delta}}\over \delta a_L(k)}\bigg).}

\vskip 1truecm
%%%%%%%%%%%%%%%%%%%%%%%%%%%

\subsec{The Wigner Functional for the Electromagnetic Field}

Now we are in a position to consider the quantum version of the
Gauss law (4.18). It is well known (see for example Ref. 29)  that the
operator equation 
$\partial_j \hat{E}_j(x) = 0$ is inconsistent with commutation 
relations (4.1) or (4.4). To avoid this inconsistency one imposes 
the weaker constraint on the ``physical states''

\eqn\ciendos{ \partial_j \hat{E}_j(x) |\Psi^{phys} \rangle = 0. }
which is equivalent to

\eqn\cientres{
\bigg( \hat{a}_L(k) + \hat{a}^{^{\dagger}}_L(-k) \bigg)
|\Psi^{phys}\rangle = 0 }
or in terms of $\hat{Q}_L$ and $\hat{P}_L$

\eqn\ciencuatro{ \bigg( \hat{Q}_L(k) + \hat{Q}_L(-k)\bigg) |\Psi^{phys} \rangle =0 \ \ \ and \ \ \
\bigg( \hat{P}_L(k) - \hat{P}_L(-k)  \bigg) | \Psi^{phys} \rangle =0. 
}

The Wigner functional in the case of electromagnetic field is defined
similarly as for the scalar field. Let $\hat{\rho}^{phys}$ be the density
operator of a physical quantum state of the electromagnetic field. Then
the Wigner functional corresponding to this state is given by

\eqn\ciencinco{
\rho_{_W}[Q,P] = \int {\cal D} ({\xi \over 2 \pi \hbar}) exp \bigg\{ - {i
\over \hbar} \int dk \xi(k) P(k) \bigg\} \langle Q + {\xi\over 2} |
\hat{\rho}^{phys}| Q - {\xi \over 2} \rangle \propto Tr \bigg\{
\hat{\Omega}[Q,P] \hat{\rho}^{phys}\bigg\}. }

When $\hat{\rho}^{phys} = |\Psi^{phys} \rangle \langle \Psi^{phys}|$ then Eq. \ciencinco\
gives

\eqn\cienseis{
 \rho_{_W}[Q,P] = \int {\cal D} ({\xi \over 2 \pi
\hbar}) exp \bigg\{ - {i \over \hbar} \int dk \xi(k) P(k) \bigg\} \Psi^{phys*}[Q -
{\xi \over 2}] \Psi^{phys}[Q + {\xi \over 2}]. }

In deformation quantization formalism Eqs. (4.30) or (4.31)
read

\eqn\ciensiete{
\bigg({a}_L(k) + {a}^*_L(-k) \bigg) * \rho_{_W}[a,a^*] = 0 }
or
\eqn\cienocho{ \bigg( {Q}_L(k) + {Q}_L(-k) \bigg) * \rho_{_W}[Q,P] = 0 \ \ \ {\rm and}
\ \ \ 
\bigg( {P}_L(k) - {P}_L(-k) \bigg)* \rho_{_W}[Q,P] =0,}
respectively. Using (4.28) one gets

$$ \bigg\{{Q}_L(k) + {Q}_L(-k) + {i\hbar \over 2} \bigg( {\delta
\over \delta P_L(k)} + {\delta  \over \delta P_L(-k)}\bigg)
\bigg\}\rho_{_W}[Q,P] =0, $$

\eqn\ciennueve{ \bigg\{{P}_L(k) - {P}_L(-k) - { i\hbar  \over 2} \bigg(
{\delta  \over \delta Q_L(k)} - {\delta \over \delta
Q_L(-k)}\bigg) \bigg \} \rho_{_W}[Q,P] = 0. }
Employing the fact that the Wigner functional is real we obtain four
equations

$$ \bigg(Q_L(k) + Q_L(-k)\bigg) \rho_{_W}[Q,P] =0, \ \ \  
\bigg({P}_L(k) - {P}_L(-k) \bigg) \rho_{_W}[Q,P] =0
$$
\eqn\otrascond{
\bigg( {\delta
\over \delta Q_L(k)} - {\delta  \over \delta Q_L(-k)}\bigg)
\rho_{_W}[Q,P] =0, \ \ \  \bigg(
{\delta  \over \delta P_L(k)} + {\delta \over \delta
P_L(-k)}\bigg)  \rho_{_W}[Q,P] = 0. }
The general solution of these equations is given by

\eqn\gensol{
\rho_{_W}[Q,P] = \rho^T_{_{W}}[Q_T,P_T] \delta[Q_L(k) + Q_L(-k)] \delta[P_L(k) - P_L(-k)].}
Comparing this result with the formula for classical constraints \constra\ one observs that
the Wigner functional $\rho_{_W}$ vanishes on the points of phase space ${\cal Z}_E$ which
don't satisfy these constraints. (Compare with Ref. 10).

\vskip 1truecm
%%%%%%%%%%%%%%%%%%%%%%%%
\noindent
{\it Example: The Ground State}

The Wigner functional $\rho_{_{W0}}$ of the ground state is defined by

$$ \bigg( {a}_L(k) + \hat{a}_L^{*}(-k) \bigg) * \rho_{_{W0}} = 0 $$

\eqn\ciendiez{ {a}_{T}(k) * \rho_{_{W0}} = 0.}
By Eq. \gensol\ we are led to the equation 

\eqn\trans{ {a}_{T}(k) * \rho^T_{_{W0}} = 0.}
Employing (4.28) we have
\eqn\vacua{
a_T \rho^T_{_{W0}} + {1 \over 2} {\delta \rho^T_{_{W0}} \over \delta a_T^*(k)} = 0.}
Thus the ground state is given by

\eqn\sol{
\rho^T_{_{W0}} = C exp \bigg( -2 \int dk \ a^*_T(k) a_T(k) \bigg), \ \ C >0.}
Hence the Wigner functional (4.38) for the ground state reads

\eqn\cienquince{
\rho_{_{W0}}[Q,P]= C
exp \bigg\{ -{1 \over \hbar}
\int {dk \over \omega (k)} \bigg( P_T^{2}(k) + \omega ^2 (k) Q_T^2 (k)
\bigg) \bigg \}  \delta[Q_L(k) + Q_L(-k)] \delta[P_L(k) - P_L(-k)].}

Similarly as in the case of scalar field we can find the Wigner functional
for any higher state. To this end one must change in the formula (3.27)
$a^*$ and $a$ by $a^*_{T \alpha}$ and $a_{T\alpha}$. 

Let
$\hat{\cal O}$ be a gauge invariant quantum observable and let
$\hat{\rho}^{phys}$ be the density operator of the physical state. The
action of $\hat{\cal O}$ on $\hat{\rho}^{phys}$ is equivalent to the action
of transversal part $\hat{\cal O}_T$ of $\hat{\cal O}$ on
$\hat{\rho}^{phys}$.  Let ${\cal O}_T[Q_T,P_T]$ be the functional
corresponding to $\hat{\cal O}_T$, ${\cal O}_T[Q_T,P_T]= W^{-1} \big(
\hat{\cal O}_T \big)$. Then for the expected value $\langle \hat{\cal O}
\rangle$ one gets

$$
\langle \hat{\cal O} \rangle = {Tr \{ \hat{\cal O} \hat{\rho}^{phys} \}  \over
Tr \{ \hat{\rho}^{phys} \} } $$

\eqn\ciendseis{ = 
{\int {\cal D}Q_T {\cal D} ({P_T \over 2 \pi \hbar}) {\cal O}_T[Q_T,P_T]
{\rho}^T_{_W}[Q_T,P_T] \over \int {\cal
D}Q_T {\cal D} ({P_T \over 2
\pi \hbar}){\rho}^T_{_W}[Q_T,P_T]}. }

\vskip 1truecm
%%%%%%%%%%%%%%%%%%%%%%%%%%%

\subsec{Ordering }

Finally we can also define the normal ordering of field operators
within deformation quantization formalism. It can be done, as before,
with the use of the operator $\hat{\cal N}_T$ acting in the phase space
${\cal Z}_E$

$$
\hat{\cal N}_T := exp \bigg \{ - {\hbar \over 4} \int dk \bigg( \omega(k)
{ \delta^2 \over \delta P_T^2(k)} + {1 \over \omega(k)}{\delta^2 \over
\delta Q_T^2(k)} \bigg) \bigg\} $$

\eqn\ciendsiete{ = exp \bigg\{ - {1\over 2} \int dk { \delta^2 \over
\delta
a_T(k) \delta a_T^*(k)} \bigg \}.}

Let ${\cal O}[Q,P]$ be any gauge invariant functional on ${\cal Z}_E$
and $\hat{\cal O}$ its Weyl image $\hat{\cal O} = W \big( {\cal
O}[P,Q]\big)$. Then, as before

\eqn\ciendocho{ : \hat{\cal O}: \  \buildrel{df}\over {=} W \bigg(
\hat{\cal N}_T {\cal O}[Q,P] \bigg). }

Having done all that one can easily formulate the deformation quantization of
electromagnetic field in the Coulomb gauge: $A_4 = 0$ and $\partial_jA_j=0$.
Here$^{29}$ $\{A_i(x,t),E_j(y,t)\} = -\delta_{ij} \delta^{T}(x-y)$, where $\delta^T$
stands for the transversal $\delta$-function. Consequently, from the very begining 
$A=(A_1,A_2,A_3)$ and  $-E=-(E_1,E_2,E_3)$ are no longer independent canonical 
variables. However, $Q_{T\alpha}$ and $P_{T\alpha}$ are such variables. Moreover, 
the constraint $\partial_jE_j(x)=0$ is automatically satisfied. Therefore 
results obtained for the temporal gauge can be quickly carry over to the Coulomb 
gauge by omitting the longitudinal parts in all formulas of temporal gauge.

\vskip 2truecm
%%%%%%%%%%%%%%%%%%%%%%%%%%%%%%%%%%%%%%%%%%%%%%%%%%%%%%%%%%%%%%%%%%%%%%
%%%%%%%%%%%%%%%%%%%%%%%%%%%%%%%%%%%%%%%%%%%%%%%%%%%%%%%%%%%%%%%%%%%%%%

\newsec{Topological Effects (Casimir Effect) in Deformation Quantization}

In this section we are going to compute the vacuum expectation value of
the energy of a real scalar field on the cylinder and on the M\"obius strip 
(twisted scalar field) within the deformation quantization formalism.   

%%%%%%%%%%%%%%%%%%%%%%%%%%%%%%%%%%%%%%%%%%%%%%%%%%%%%%%%%%%%
\subsec{Scalar Field on the Cylinder}

Consider a cylinder ${\bf S}^1 \times \IR$ representing a two dimensional
space time. Here ${\bf S}^1$ is the spatial part and $\IR$ is the temporal
part. Local coordinates are as usual $(x,t)$. ${\bf S}^1$ has
circumference $L$, then $k$ is quantized as $k = {2 \pi \over L} n $ with
$n \in \IZ$ and the frequency is given by $\omega(k) = \sqrt{ k^2 + m^2} =
\sqrt{ {4\pi^2 \over L^2} n^2 + m^2}$. The Hamiltonian operator can be
seen as the zero-zero component of the energy-momentum tensor

\eqn\ciendnueve{
 \hat{T}_{00}= {1\over 2} \bigg( (\partial_t \hat{\phi})^2 +
(\partial_x \hat{\phi})^2 + m^2 \hat{\phi}^2 \bigg).}

Now we would like to compute the vacuum expectation value $\langle
\hat{T}_{00} \rangle (L)$. In order to do this computation we will use the
point splitting method$^{35,36}$ and we write

$$
\langle \hat{T}_{00} \rangle (L) = \lim_{t \to t'}\lim_{x \to x'} 
\bigg\{ \langle 0_L| {1 \over 2}
\bigg(\partial_t \partial_{t'} + \partial_x \partial_{x'} + m^2 \bigg)
\hat{\phi}(x,t)\hat{\phi}(x',t')|0_L \rangle
$$

\eqn\cienveinte{
 - \langle 0_{\infty} | {1 \over 2}
\bigg( \partial_t \partial_{t'} + \partial_x \partial_{x'} + m^2 \bigg)
\hat{\phi}(x,t) \hat{\phi}(x',t') | 0_{\infty} \rangle \bigg\}, }
where $|O_L \rangle$ and $|O_{\infty} \rangle$ are the vacuum states for
the 
two dimensional cylindrical
and Minkowski space times, respectively. 

From the fact that the second term of the right-hand side of the above
equation is independent of $L$ we can rewrite it as follows

\eqn\cienvuno{ \langle \hat{T}_{00} \rangle (L):= \int dL \ 
\lim_{t \to t'} \lim_{x \to x'} \bigg\{ {1 \over 2}
\bigg(\partial_t
\partial_{t'} + \partial_x \partial_{x'} + m^2 \bigg) {\partial \over
\partial L} \langle 0_L |
\hat{\phi}(x,t)\hat{\phi}(x',t')|0_L \rangle \bigg\} }
where the integration constant is defined by the condition 
$\langle \hat{T}_{00} \rangle (\infty) =0$.

Thus to compute of $\langle \hat{T}_{00} \rangle (L)$ it is necessary first to
compute the quantity $\langle 0_L | \hat{\phi}(x,t)$ $\hat{\phi}(x',t')|0_L \rangle$.
In terms of deformation quantization we have (compare with (4.44))

\eqn\cienvdos{ \langle 0_L | \hat{\phi}(x,t) \hat{\phi}(x',t') | 0_L \rangle =
 {\int {\cal D}Q {\cal D}
({P \over 2 \pi \hbar}) \phi (x,t) * \phi (x',t') \rho^L_{W_0} [Q,P] 
\over \int {\cal D}Q {\cal D}({P \over 2 \pi \hbar}) \rho^L_{W_O}[Q,P] }, }
where $\rho^L_{W_0}[Q,P]$ is the Wigner functional of the ground state 
(see (3.21))

\eqn\cienvtres{ \rho^L_{W_0} \propto exp \bigg\{ - {1 \over \hbar} \sum_k 
{1 \over \omega(k)}\bigg(P^2(k) + \omega^2(k)Q^2(k)\bigg)\bigg\} }
and (see (3.11))

\eqn\cienvcuatro{ \phi (x,t)= { 1 \over \sqrt{L} } 
\sum_k \bigg( Q(k) \cos \big(kx-\omega(k)t\big) - {1 \over \omega 
(k)} P(k) \sin\big(kx -\omega(k)t\big) \bigg). }

After straightforward calculations we get

\eqn\cienvcinco{ \langle 0_L |
\hat{\phi}(x,t)\hat{\phi}(x',t')|0_L \rangle = {\hbar \over 2L} \sum_k
{exp\big\{ i\big(k(x-x') -\omega(k)(t-t')\big) \big\} \over \omega (k)}. }

Substituting \cienvcinco\ into \cienvuno\ and using some considerations given 
by Kay$^{37}$ one finds 

\eqn\cienvseis{\langle \hat{T}_{00} \rangle(L) = \int dL \bigg\{ -{m \hbar 
\over 2L^2} + 2 \pi \hbar
\lim_{\sigma \to 0} \partial_L \bigg[{ 1 \over L^2} \bigg(S(a) + O(z) - 
{1 \over 4} \big({1   
\over \sin^2(z/2)} \big) - {a^2 \over 2} \ln \big( 2 \sin (z/2)\big) \bigg)  
\bigg] \bigg\}  }
where $z = {2 \pi \over L}\sigma$, $a={mL \over 2 \pi}$ and 

$$
S(a) = \sum_{n=1}^{\infty} \bigg(\sqrt{n^2 + a^2} - n - 
{a^2 \over 2n}\bigg), 
$$
\eqn\cienvsiete{
O(z) = \sum_{n=1} \bigg(\sqrt{n^2 + a^2} - n - {a^2 \over 2n} \bigg) 
\bigg(\cos(nz)-1 \bigg). }
Hence

\eqn\cienvocho{ \langle \hat{T}_{00} \rangle (L) = 2 \pi \hbar \bigg[- {1 \over 12
L^2} + {m \over 4 \pi L} + {S(a) \over L^2} + { m^2 \over 8 \pi^2} 
\ln(mL) \bigg] + D, }
where $D$ is the integration constant which can be computed by the condition $\lim_{L \to \infty}
\langle \hat{T}_{00} \rangle (L) = 0$. Thus 
   
$$ D = - 2 \pi \hbar \lim_{L \to \infty} \bigg( {S(a) \over L^2} +
{m^2 \over 8 \pi^2} \ln (mL) \bigg) = -{m^2 \hbar \over 2 \pi} \lim_{a \to \infty} \bigg(
{1 \over a^2} {\cal F}(a^2) \bigg) - {m^2 \hbar \over 4 \pi} \ln (2 \pi) $$

\eqn\cienvnueve{ {\cal F}(a^2) = \sum_{n=1}^{\infty} \bigg( \sqrt{n^2 +
a^2} -  n  -{a^2 \over 2n}\bigg) +{a^2 \over 4} \ln a^2. }

Employing the relation involving the $K_0$ Bessel function$^{39}$

\eqn\cientreinta{ \sum_{n=1}^{\infty} K_0(nx) =
{1\over 2}( {\bf C} + \ln {x \over 4 \pi}) +  {\pi \over 2x} + \pi
 \sum_{l=1}^{\infty} \bigg( {1 \over \sqrt{x^2 + 4 \pi^2 l^2}}  - {1 \over
2 \pi l} \bigg), }
where ${\bf C}$ is Euler's constant and setting $x = 2 \pi a$ we find the
following relation

\eqn\cientuno{ {\partial {\cal F}(a^2) \over \partial a^2} =
\sum_{n=1}^{\infty}K_0(2 \pi n a) - {1 \over 2} {\bf C} + {1\over 2}
\ln 2 - {1 \over 4 a}+ {1\over 4}. }

Integrating \cientuno\ with respect to $da^2$ and dividing the result by 
$a^2$ one finally gets

\eqn\cientdos{ \lim_{a \to \infty}{1 \over a^2}{\cal F}(a^2) = - {1 \over 2}\bigg(
{\bf C} - \ln 2 - {1 \over 2} \bigg). }

Substituting \cientdos\ into \cienvnueve\ we get the integration constant to
be
  
\eqn\cienttres{ D = {m^2 \hbar \over 4 \pi} \bigg({\bf C} - {1 \over 2} - 
\ln 4 \pi \bigg). }

The formulas \cienvocho\ with \cienttres\ give the final result which for standard quantum field theory has
been obtained by Kay$^{37}$.

For the massless case $(m=0)$ we get$^{37,38,40}$

\eqn\cientcuatro{\lim_{m \to 0} \langle \hat{T}_{00} \rangle (L) = - 
{\pi \hbar \over 6 L^2}.}

\vskip 1truecm
%%%%%%%%%%%%%%%%%%%%%%%%%%%%%%%%%%%%%%%%%%%%%%%%%%%%%%%%%%%%
\subsec{Scalar Field on The M\"obius Strip}

Here we deal with the case of scalar field on the M\"obius strip$^{38,40}$.
Now the quantization rule gives $k={2 \pi (n + {1 \over 2}) \over L}$. Analogous calculations as in
the previous case lead to the result 

\eqn\cientcinco{ \langle \hat{T}_{00} \rangle (L) = 2 \pi \hbar \int dL 
\lim_{\sigma \to 0} \partial_L
\bigg[ {1 \over L^2} \bigg( S'(a) + O'(z) - {1 \over 4} {\cos (z/2) \over 
\sin^2(z/2)} + {a^2 \over 2}
\ln[ \cot(z/4)] \bigg) \bigg], }
where $z= {2 \pi \over L} \sigma$, $a={mL \over 2 \pi}$ and

$$ S'(a) = \sum_{n=1}^{\infty} \bigg( \sqrt{(n - {1\over 2})^2 + a^2} - (n -  
{1\over 2}) - {a^2 \over 2(n - {1\over 2})} \bigg), $$

\eqn\cientseis{ O'(z) = \sum_{n=1}^{\infty} \bigg(\sqrt{(n - {1\over 2})^2 + a^2} -
(n - {1 \over 2}) - {a^2 \over
2(n - {1\over 2})} \bigg) \bigg( \cos[(n - {1\over 2})z]-1 \bigg). }
Hence 

\eqn\cientsiete{\langle \hat{T}_{00} \rangle (L) =  2 \pi \hbar 
\bigg[ {S'(a) \over L^2} + {1 \over 24 L^2}
+ {m^2 \over 8 \pi^2} \ln(mL)\bigg] + D'. }

The integration constant can be computed from the condition
$ \lim_{L \to \infty} \langle \hat{T}_{00} \rangle (L) = 0 $
and it yields

$$ D' = - 2 \pi \hbar  \lim_{L \to \infty} \bigg( {S'(a) \over L^2} +
{m^2 \over 8 \pi^2} \ln (mL) \bigg)= -{m^2 \hbar \over 2 \pi} \lim_{a \to \infty} \bigg( 
{1 \over a^2} {\cal F'}(a^2) \bigg) - {m^2 \hbar \over 4 \pi} \ln (2 \pi),
$$

\eqn\cientocho{ {\cal F'}(a^2) = \sum_{n=1}^{\infty} \bigg( \sqrt{(n - {1 \over 2})^2 +
a^2} - ( n - {1\over 2}) -{a^2 \over 2(n - {1\over 2})}\bigg) +{a^2 \over
4} \ln a^2.  }

Now using the following relation$^{39}$

\eqn\cientnueve{ \sum_{n=1}^{\infty} (-1)^n K_0(nx) =
{1\over 2}( {\bf C} + \ln {x \over 4 \pi}) + \sum_{l=1}^{\infty}
\bigg( {1 \over \sqrt{(2l -1)^2 + ({x\over \pi})^2}} - {1 \over 2l}
\bigg)}
and setting $x=2\pi a$ we find

\eqn\ciencuarenta{ {\partial {\cal F'}(a^2) \over \partial a^2} =
\sum_{n=1}^{\infty}(-1)^n K_0(2 \pi n a) - {1 \over 2} {\bf C} - {1\over 2}
\ln 2 + {1\over 4}. }

Integrating \ciencuarenta\ with respect to $da^2$ and dividing by $a^2$ one obtains

\eqn\ciencuno{ \lim_{a \to \infty}{1 \over a^2}{\cal F'}(a^2) = - {1 \over 2}
\bigg( {\bf C} + \ln 2 - {1 \over 2} \bigg). }

Substituting \ciencuno\ into \cientocho\ we have 

\eqn\ciencdos {D' = {m^2 \hbar \over 4 \pi} \bigg( {\bf C} - {1 \over 2} - 
\ln \pi\bigg).}

Finally, for the twisted scalar field one gets 

\eqn\cienctres{ \langle \hat{T}_{00} \rangle (L) = 2 \pi \hbar 
\bigg(  {1 \over 24 L^2} + {S'({mL \over 2 
\pi}) \over L^2} + {m^2 \over 8 \pi^2} \ln (mL) + 
{m^2 \over 8 \pi^2} \big( {\bf C} - {1 \over 2} - \ln \pi
\big) \bigg). }

For $m=0$ we recover the result found by Isham$^{40}$ (see also Ref. 38)

\eqn\cienccuatro{ \langle \hat{T}_{00} \rangle (L) = {\hbar \pi \over 12 L^2}. }

\vskip 2truecm
%%%%%%%%%%%%%%%%%%%%%%%%
%%%%%%%%%%%%%%%%%%%%%%%

\subsec{New Normal Ordering}

We end this section with some comments which will be developed in further work.
The results obtained suggest that it seems to be reasonable (and perhaps necessary) to deal
with new normal ordering $\hat{\cal N}'$ when spaces of non-trivial topology are considered.
In the present case $\hat{\cal N}'$ should satisfy the following condition

\eqn\cienccuatro{ \langle \hat{H}' \rangle (L) = {\int {\cal D}Q {\cal D}
({P \over 2 \pi \hbar}) (\hat{\cal N}'H)\rho^L_{W_0} [Q,P] 
\over \int {\cal D}Q {\cal D}({P \over 2 \pi \hbar}) \rho^L_{W_O}[Q,P] }}
where $\langle \hat{H}' \rangle (L)$ is the vacuum energy. 
The simplest and natural assumption is (see (3.28))

\eqn\cienccinco{\hat{\cal N}' = exp \bigg( \sum_k \big(- {1 \over 2} + \gamma(k) \big)
{\partial^2 \over \partial a(k) \partial a^*(k)} \bigg).
}

Inserting (5.28) into (5.27), employing the formulas (3.31) and (3.35) one gets the condition on
$\gamma(k)$

\eqn\ciencseis{ \hbar \sum_k \gamma(k) \omega(k) = \langle \hat{H}' \rangle (L). }

Thus, for example, in the case of the massless scalar field on cylindrical space time we 
have (see (5.16))

\eqn\ciencsiete{\sum_{n \not= 0} |n| \gamma({2 \pi n \over L}) = - {1 \over 12}. }
Of course, there is infinite number of solutions to (5.29) or (5.30).

Given $\hat{\cal N}'$ one can define new star product $*'$ which is cohomologically equivalent 
to the Moyal $*$-product

\eqn\ciencocho{F_1 *' F_2 = \hat{\cal N}'^{-1}\bigg(\hat{\cal N}' F_1 * \hat{\cal N}' F_2\bigg).}
The star product $*'$ gives a new quantization of classical field.

Consider now $\lambda \phi^4$-field theory on the cylindrical space time. We have

\eqn\ciencnueve{ H = {1 \over 2} \int dx [(\varpi)^2 + (\partial_x \phi)^2 + m^2\phi^2
+ \lambda \phi^4]. }
To fulfill condition (5.27) in first order of perturbation theory we take now 

\eqn\cienci{\hat{\cal N}' = exp \bigg\{ \sum_k \bigg( \big(- {1 \over 2} + \gamma(k) \big)
{\partial^2 \over \partial a(k) \partial a^*(k)} + \nu(k) {\partial^4 \over \partial a^2(k) \partial
a^{*2}(k)}\bigg) \bigg\}. }

Straightforward calculations show that the condition (5.27) leads to the following relations

$$
\sum_k {\gamma(k) \over \omega(k)}=0, \ \ \ \ \  \sum_k {\gamma(k) \over 
\omega^2(k)}=0
$$

\eqn\cienciuno{
\hbar \sum_k \gamma(k) \omega(k) + {3 \hbar^2 \lambda \over 4 L} \sum_k {\nu(k) \over
\omega^2(k)} = \langle \hat{H} ' \rangle (L),}
where $k = {2\pi n \over L}$ and $ \omega(k) = \sqrt{ ({2 \pi n \over L})^2 + m^2}$
with $n \in \IZ.$ ($ \langle \hat{H} ' \rangle (L)$ for $\lambda \phi^4$-field theory
has been found by Kay$^{37}$).

Further developing of non-linear field theory in terms of deformation quantization
formalism is a very difficult problem as in that case we must look for other
cohomologically equivalent star products to avoid divergences (Dito$^8$). We are going
to consider this question in a separate paper.

\vskip 2truecm
%%%%%%%%%%%%%%%%%%%%%%%%%%%%%%%%%%%%%%%%%%%%%%%%%%%%%%%%%%%%%%%%%%%%%%
%%%%%%%%%%%%%%%%%%%%%%%%%%%%%%%%%%%%%%%%%%%%%%%%%%%%%%%%%%%%%%%%%%%%%%

\newsec{Final Remarks}

In this paper we have generalized some aspects of deformation quantization to 
non-interacting field theory. We were able to show that many well known results 
of deformation quantization in quantum mechanics could be extended to the case of 
quantum field theory. 
This was possible because a free field 
can be represented
as an infinite number of independent harmonic oscillators. One can apply the usual deformation
quantization formalism to each oscillator to obtain deformation quantization
of the whole theory. Consequently it is
expected that phase space interpretation of quantum field theory can be also
extended to perturbative field theory. Some work in this direction has been done 
by Dito$^8$, but many problems remain to be investigated. For example the
deformation quantization of $\sigma$-model and Chern-Simons gauge theory, which
have non-trivial phase spaces. Interesting is if Fedosov's approach$^4$ 
can be applied in the later cases. We intend to devote a forthcoming work
to these questions.

\vskip 2truecm
%%%%%%%%%%%%%%%%%%%%%%%%%%%%%%%%%%%%%%%%%%%%%%%%%%%%%%%%%%%%
\centerline{\bf Acknowledgements}

This work was partially supported by CONACyT and CINVESTAV (M\'exico). One
of us (M.P.) is indebted to the staff of Departamento de F\'{\i}sica,
CINVESTAV, M\'exico D.F., for warm hospitality.

We are grateful to Referee for valuable comments and especially for pointing out
the error in Sec. 4 of the previous version of our paper. 

%%%%%%%%%%%%%%%%%%%%%%%%%%%%%%%%%%%%%%%%%%%%%%%%%%%%%%%%%%%%%%%%%%%%%%%%%%%

%%%%%%%%%%%%%%%%%%%%%%%%%%%%%%%%%%%%%%%%%%%%%%%%%%%%%%%%%%%%%%%%%%%%%%%%%%%

\listrefs

\end